\documentclass[twocolumn]{aastex62}
\usepackage[version=4]{mhchem}
\usepackage{gensymb}
\usepackage{natbib}
\graphicspath{{./}{figures/}}

\received{November 2, 2018}
\revised{February 11, 2019}
\accepted{February 11, 2019}

\submitjournal{ApJ}

\shorttitle{Sample article}
\shortauthors{Sie et al.}

\begin{document}

\title{On The Photodesorption of \ce{CO2} Ice Analogues: The Formation of Atomic C in The Ice and the Effect of The VUV Emission Spectrume}

\correspondingauthor{Y.-J. Chen}
\email{asperchen@phy.ncu.edu.tw}

\author{N.-E. Sie}
\affil{Department of Physics, National Central University, Jhongli City, Taoyuan County 32054, Taiwan}

\author{G.M. Mu\~{n}oz Caro}
\affiliation{Centro de Astrobiolog\'ia (INTA-CSIC), Carretera de Ajalvir, km 4, Torrej\'on de Ardoz, 28850 Madrid, Spain}

\author{Z.-H. Huang}
\affiliation{Department of Physics, National Central University, Jhongli City, Taoyuan County 32054, Taiwan}

\author{R. Martín-Doménech}
\affiliation{Center for Astrophysics, Harvard \& Smithsonian, 60 Garden Street, Cambridge, MA 02138, USA}

\author{A. Fuente}
\affiliation{ Observatorio Astronómico Nacional (OAN,IGN), Apdo. 112, E-28803 Alcalá de Henares, Spain}

\author{Y.-J. Chen}
\affiliation{Department of Physics, National Central University, Jhongli City, Taoyuan County 32054, Taiwan}

\begin{abstract}
\ce{CO2} ice has a phase transition at 35 K when its structure changes from amorphous to crystalline. Using Reflection absorption Infrared Spectroscopy (RAIRS), \cite{oberg09} observed that the photodesorption yield of \ce{CO2} ice deposited at 60 K and irradiated at 18 K is 40\% lower than that of \ce{CO2} ice deposited and irradiated at 18 K. In this work, \ce{CO2} ices were deposited at 16--60 K and UV-irradiated at 16 K to rule out the temperature effect and figure out the relationship between photodesorption yield and ice structure. \\IR spectroscopy is a common method used for measurement of the photodesorption yield in ices. We found that undetectable C atoms produced in irradiated \ce{CO2} ice can account for 33\% of the amount of depleted \ce{CO2} molecules in the ice. A quantitative calibration of QMS was therefore performed to convert the measured ion current into photodesorption yield. During various irradiation periods, the dominant photodesorbing species were CO, \ce{O2}, and \ce{CO2}, and their photodesorption yields in \ce{CO2} ices deposited at different temperature configurations were almost the same, indicating that ice morphology has no effect on the photodesorption yield of \ce{CO2} ice. In addition, we found that the lower desorption yield reported by \cite{martin15} is due to a linear relationship between the photodesorption yield and the combination of energy distribution of Microwave-Discharge Hydrogen-flow Lamp (MDHL) and UV absorption cross section of ices.
\end{abstract}

\section{Introduction} \label{sec:intro}
In the interior of dense clouds, gaseous molecules collide with dust grains and accrete to form ice mantles due to the low dust grain temperature of around 10 K. Desorption of ice molecules induced by cosmic rays and secondary ultraviolet (UV) photons is expected to contribute to the observed gas phase abundances toward cold regions where thermal desorption is inhibited \citep{G16, oberg07CO, westley1995, willacy1998}. The secondary UV field is produced by the interaction of cosmic rays and hydrogen molecules \citep{cecchi1992, weinberg2003emergence, shen2004cosmic}. In general, when a molecule in the ice bulk absorbs a photon, it becomes excited, and may transfer a fraction of its energy to the molecules near the surface leading to their desorption, which is the so called desorption induced by electronic transition (DIET) \citep{bertin12, fillion14, martin15}. \\\ce{CO2} is one of the most common species in interstellar ices and has been observed toward many high- and low-mass young stellar objects (YSOs), with abundances up to 19\%--28\% relative to \ce{H2O} (\citealt{boogert2015observations} and references therein). Solid \ce{CO2} is not only a significant chemical tracer of carbon and oxygen in star-forming regions \citep{van2006infrared, van2006photoprocesses, cooke2016co2}, but is an indicator of the temperature history as well \citep{kim2012co2, pontoppidan2008c2d}.\\\ce{CO2} photodesorption has already been the subject of other scientific studies, but its connection with the ice structure is still not well understood. \citep{yuan2014radiation} discussed the photodesorption rate of \ce{CO2} ice under radiation damage, indicating that the photodesorption yield of \ce{CO2} ice depends on its structure. In addition, the amorphous \ce{CO2} ice provides facile pathways for CO and \ce{CO2} molecules passing through the ice bulk and reaching the surface, which contributes to a more efficient desorption compared to crystalline \ce{CO2} ice \citep{cooke2018co}. \citep{bahr2012photodesorption} measured the photodesorption yields of \ce{CO2} irradiated by Ly$\alpha$ photons at 16--60 K, showing that the photodesorption yield of \ce{CO2} ice strongly depends on temperature. However, CO molecules are efficiently produced in the ice bulk due to \ce{CO2} photodissociation during irradiation. Thermal desorption of photo-produced CO molecules near 30 K may influence the photodesorption of \ce{CO2} above this temperature. In \cite{oberg09}, the \ce{CO2} deposited at 60 K and cooled down to irradiate at 18 K has a 40\% lower photodesorption yield compared to the \ce{CO2} deposited at 18 K, suggesting that the \ce{CO2} photodesorption yield depends on temperature as well as structure. On the other hand, the photodesorption yield of CO ice is found to be linearly temperature dependent between 8--20 K in \cite{G16}, while the structure of CO ice changed only above 20 K, which means that the photodesorption yield of CO is not directly related to the CO ice structure. In addition to a more efficient photodissociation of \ce{CO2}, compared to CO ice, the former is an apolar molecule and the later has a weak dipole moment. \\In order to investigate the relationship between the photodesorption yield of \ce{CO2} ice and its morphology, \ce{CO2} ice was deposited at 16 and 30 K to produce amorphous structure whereas deposition at 40, 50, and 60 K served to create crystalline \ce{CO2} ice, since the phase change occurs at $\sim$ 30 K \citep{escribano2013crystallization}. After deposition, the vacuum-ultraviolet (VUV) irradiation of \ce{CO2} ice was done at 16 K to measure the photodesorption yield in the absence of thermal desorption while preserving the initial ice structure. According to \cite{escribano2013crystallization}, the morphology of \ce{CO2} ice depends on temperature, which can provide us the temperature history of the environment \citep{d1989,kim2012co2,pontoppidan2008c2d}. In this report, we executed a quantitative calibration of a quadrupole mass spectrometer (QMS) to convert the relative ion current value to a photodesorption yield. Differences in the photodesorption yields between this work and \cite{martin15} come from the weighted-absorption intensity from the different emission spectra of the VUV lamps, which means the photodesorption yield is positively proportional to the weighted-absorption intensity.
\section{Experimental Methods} \label{sec:exp}
\subsection{Experimental procedure}\label{sec:exppro}
The experiments were performed using the Interstellar Photoprocess System (IPS), which includes three parts: the main system, the gas-line and detectors \cite{chen14}. The main system is an ultra-high-vacuum (UHV) chamber with background pressure $\approx 3\times 10^{-10}$ torr, equipped with a closed-cycle helium cryostat to cool down the KBr substrate to 16 K. The gas-line is used for preparing \ce{CO2} gas (Matheson Tri-gas, 99.995\%) for deposition. The detectors consist of a transmission Fourier transform infrared spectrometer (FTIR) (ABB, FTLA 2000-104) and a quadrupole mass spectrometer (QMS) (MKS Instruments, Microvision 2), which are used to monitor the ice and the desorbing molecules during irradiation, respectively. The column densities of \ce{CO2} ices were controlled to be 140 $\pm$ 2 ML (1 ML = 1015 molecules cm$^{-2}$). The IR absorption strength adopted for solid \ce{CO2} is $7.6\times 10^{-17}$ cm molecule$^{-1}$ at 2342 cm$^{-1}$ \citep{bouilloud2015bibliographic, yamada1964}, and $1.1\times 10^{-17}$ cm molecule$^{-1}$ at 2138 cm$^{-1}$ for solid CO ice \citep{jiang1975}.\\The energetic VUV photons are generated by the T-type MDHL with an \ce{H2} pressure of 0.4 torr, photon energy from 6.89 to 10.88 eV (180--114 nm), $\approx 4.8\times 10^{13}$ photon cm$^{-2}$ s$^{-1}$ flux, and 45$^{\circ}$ incident angle to the substrate. The spectrum of this Microwave-Discharge Hydrogen-flow Lamp (MDHL) has been studied in detail by \cite{chen14} and is similar to that of the calculated UV field in dense cloud interiors \citep{gredel1989}.  After cooling down the system to a specific temperature, \ce{CO2} ice was deposited with the pressure at $5\times 10^{-9}$ torr for $\sim$ 4 minutes, accompanied by the collection of IR spectra in situ. The VUV irradiation was done at 16 K, and the total irradiation time was 90.5 minutes. During 12 irradiation cycles, the idle intervals were 3 minutes between cycles.
\subsection{Quantitative calibration of mass spectrometer}
CO ice was chosen for quantitative calibration of QMS. The most important characteristic of CO ice is that the conversion efficiency of CO into \ce{CO2} (the only observed photoproduct is \ce{CO2} for a CO ice thickness less than 30 ML) is less than 5\% and, in addition, the CO ice photodesorption yield is relatively high \citep{G10}. CO ice is, therefore, the best candidate for quantitative calibration of QMS. The experimental procedure of CO ice irradiation is the same as described in Section~\ref{sec:exppro}, with an ice thickness of 21 ML and VUV irradiation at 16 K for 1 hour.\\In order to correlate the relative signal of QMS ion current measured during photodesorption and the decrease of the CO ice column density, the quantification factor $k_{\rm IPS}(CO)$ is defined as
\begin{equation}\label{e1}
k_{\rm IPS}(CO)=\frac{N(\rm CO)}{A(\rm CO)}
\end{equation}
where $N(CO)$ is the column density of the photodesorbed CO derived from IR spectra, and $A(CO)$ is the accumulated ion current signal of CO during VUV irradiation. The subscript “IPS” is used to distinguish the quantification factor $k_{\rm ISAC}(CO)$ defined by \cite{martin15}, which is reciprocal to the definition of $k_{\rm IPS}(CO)$ and conducted within InterStellar Astrochemistry Chamber (ISAC). According to \cite{martin15}, the photodesorption of a species can be obtained by comparing to CO and quantified by $k_{\rm IPS}(CO)$ with the equation:
\begin{equation}\label{e2}
\begin{aligned}
N(\rm mol)=A(\rm mol)\times k_{\rm IPS}(\rm CO)
\times \frac{\sigma^+(\rm CO)}{\sigma^+(\rm mol)}\times \\\frac{I_F(\rm CO^+)}{I_F(z)}\times \frac{F_F(28)}{F_F(m)}\times \frac{S(28)}{S(m/z)}
\end{aligned}
\end{equation}
where $N(\rm mol)$ is the number of a certain desorbing molecule or atom, $A(\rm mol)$ is the accumulated ion current from QMS, $\sigma^+(\rm mol)$ the ionization cross section for first ionization of the molecule under incident electron energy in QMS, $I_F(z)$ the ionization factor, $F_F(m)$ the fragmentation factor, and $S(m/z)$ is the sensitivity of mass fragment $m/z$ to QMS. The procedure to determine this ratio of sensitivity can be obtained by the ratio of $k_{\rm QMS}\times S(m/z)$, where $k_{\rm QMS}$ is the proportionality constant, which is independent on species. The procedure of figuring out this ratio is presented in detail in \cite{martin15}, and the corresponding equation is
\begin{equation}\label{e3}
k_{\rm QMS}\times S(m/z)=0.14\times e^{\frac{-m/z}{18.65}}
\end{equation}
\section{Results and Discussion} \label{sec:res}
\subsection{VUV irradiation for \ce{CO2} ice}
\subsubsection{Photodesorption as seen by the Infrared spectrometer}
\begin{figure}
\centering
\includegraphics[width=8.5cm]{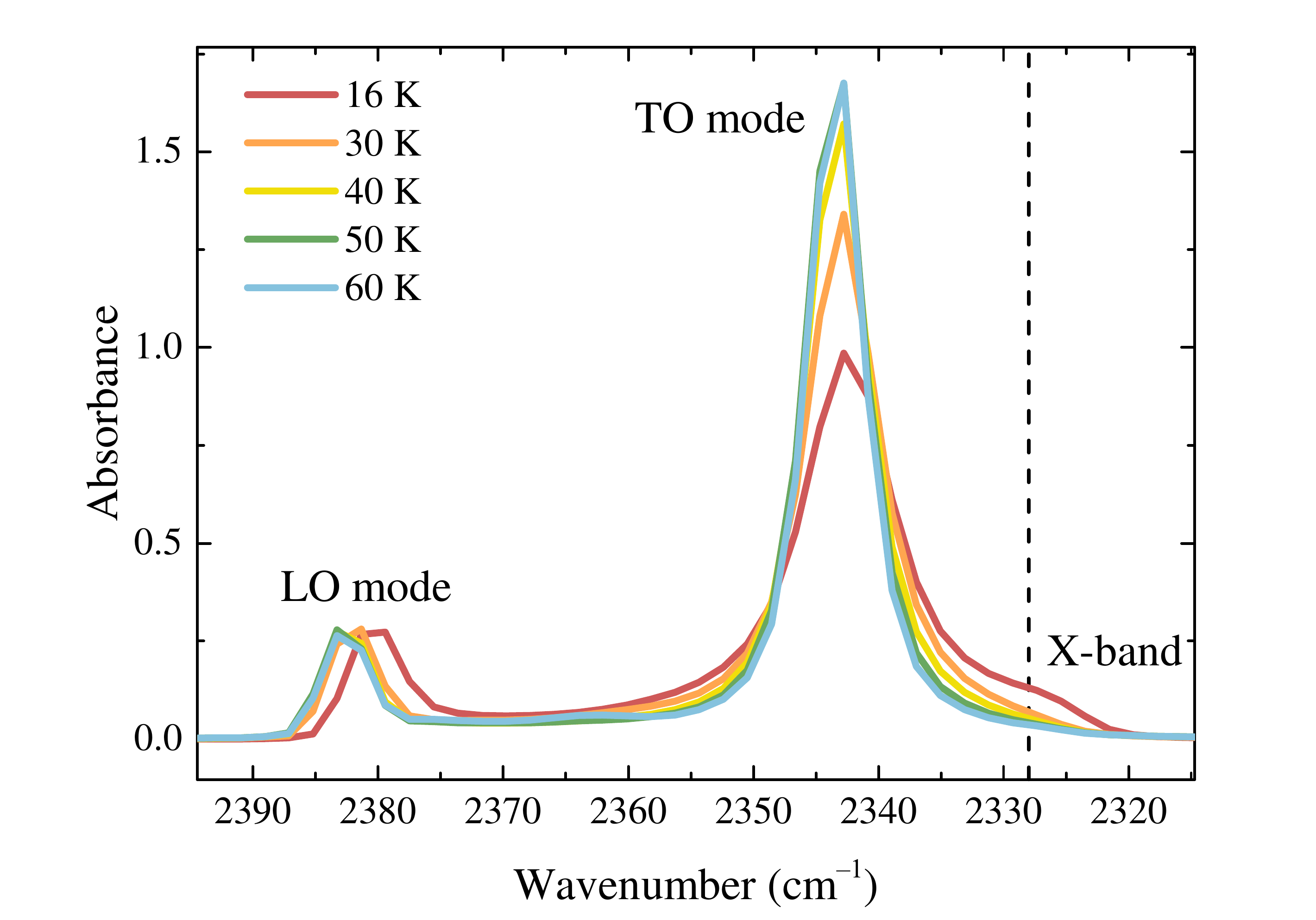}
\caption{IR absorbances of \ce{CO2} ices at different deposition temperatures. In these experiments, the initial column density of \ce{CO2} was about 140 ML.}
\label{F1}
\end{figure}

\begin{figure}
\centering
\includegraphics[width=8.5cm]{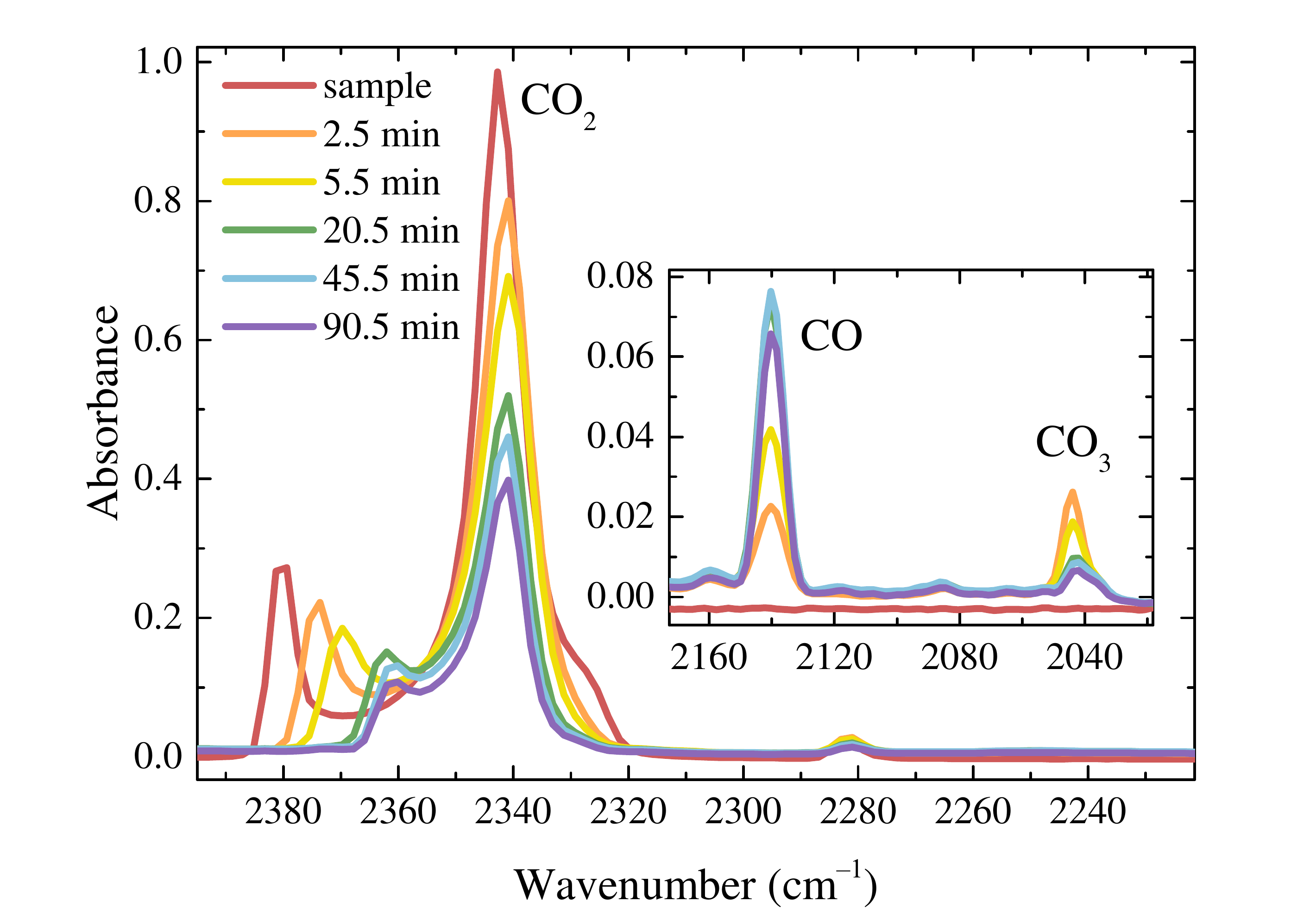}
\caption{Evolution of the absorbances of \ce{CO2} ($\sim$2340 cm$^{-1}$), CO ($\sim$2140 cm$^{-1}$), and \ce{CO3} ($\sim$2044 cm$^{-1}$) during VUV irradiation at 16 K, with the \ce{CO2} ice deposited at 16 K.}
\label{F2}
\end{figure}

From IR spectra of \ce{CO2} ices deposited at 16, 30, 40, 50, and 60 K (Figure~\ref{F1}), it is obvious that the structure of \ce{CO2} ice is temperature dependent. From the temperature programmed desorption of pure \ce{CO2} ice, the absorbance of \ce{CO2} changed at $\sim$30 K, which represents the transition to a less amorphous and porous ice \citep{escribano2013crystallization}. Hence, \ce{CO2} deposited at 16 and 30 K possesses a partially amorphous structure, whereas it is more ordered and compact at 40, 50, and 60 K configurations. The stretching band ($\nu$ 3) of \ce{CO2} is composed of longitudinal optical modes (LO mode, $\sim$2380 cm$^{-1}$) and transverse optical (TO mode, $\sim$2344 cm$^{-1}$) since the incident angle of IR beam light is 45 degrees \citep{ovchinnikov1993}. The band at $\sim$2328 cm$^{-1}$ (usually named X-band) corresponds to a pure and amorphous \ce{CO2} ice structure, which should gradually diminish during irradiation due to the formation of CO and disappear at 30 K during the annealing \citep{escribano2013crystallization}. The photochemical products can be observed in IR spectra shown in Figure~\ref{F2}. The bands at $\sim$2138 cm$^{-1}$ and $\sim$2044 cm$^{-1}$ correspond to CO and \ce{CO3}, respectively \citep{yamada1964}.\\The carbon balance method
\begin{equation}\label{e4}
\rm |\Delta CO_{2(s)}|=\Delta CO_{(s)}+\Delta CO_{3(s)}+\Delta CO_{(g)}+\Delta CO_{2(g)}
\end{equation}
is used to calculate the photodesorption yield of \ce{CO2} ice in IR measurement. The decreasing column density of \ce{CO2} ice should be converted into CO and \ce{CO3} in solid phase, along with the desorbing CO and \ce{CO2} gas phase molecules which can be regarded as photodesorption products. Hence, the total photodesorption yield can be calculated as $\rm |\Delta CO_{2(s)}|-\Delta CO_{(s)}$, neglecting the contribution of \ce{CO3} ice since it reached its largest value after 5.5 min of irradiation and became negligible in the later irradiation cycles.
\subsubsection{Photodesorption as seen by Mass spectrometer}
\begin{figure}
\centering
\includegraphics[width=8.5cm]{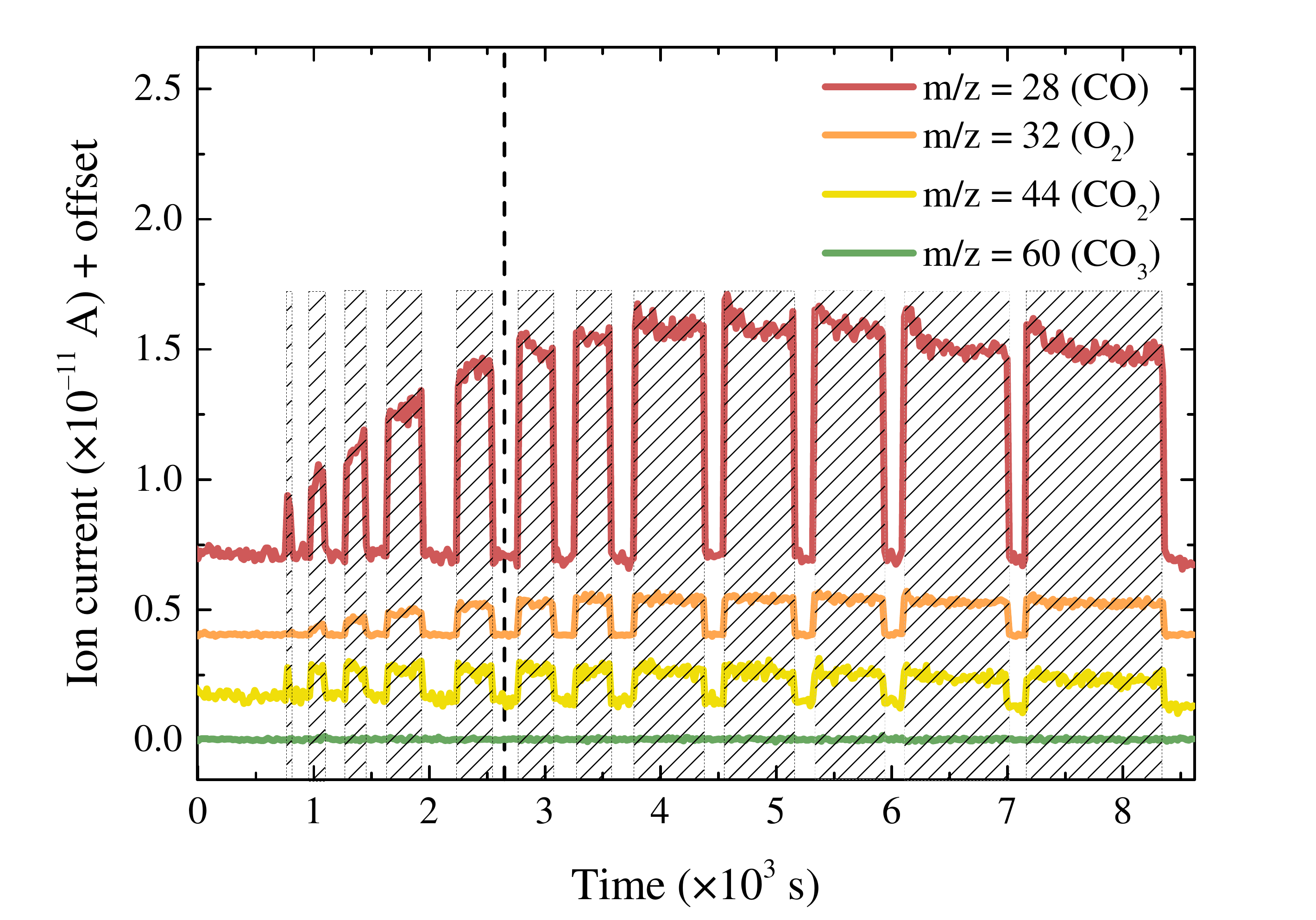}
\caption{The intensity variation of $m/z$ = 28 (CO), 32 (\ce{O2}), 44 (\ce{CO2}), and 60 (\ce{CO3}) during VUV irradiation periods. The dash line represents the photon dose reached $5\times 10^{16}$ photons cm$^{-2}$.}
\label{F3}
\end{figure}

The photon-induced desorbing species measured by QMS during VUV irradiation are shown in Figure~\ref{F3}. Although QMS is a direct method to measure desorbing species, the measured ion current does not equal to the total number of desorbing molecules because only a small fraction reaches the QMS. The quantitative values of photodesorption yields from \ce{CO2} deposited at different temperatures can be derived from calibrated QMS described in Section~\ref{sec:exppro}. After calibration, the integrated signal of each molecule represents the relative photodesorption yield of that given species. The main photo-products are $m/z$= 28 (CO), $m/z$ = 32 (\ce{O2}), and $m/z$ = 44 (\ce{CO2}), while no signal for $m/z$ = 60 (\ce{CO3}) was detected, which is consistent with the result of IR spectra. 
\\Besides the ion current during irradiation, the background signal from ambient environment is also required, which means irradiating the KBr substrate directly before deposition of \ce{CO2} ice. In addition, since gaseous \ce{CO2} is easily dissociated into CO by the ionizer of QMS giving $m/z$ = 28, we have to extract the mass spectrum of \ce{CO2} measured during deposition. An excess in the relative intensity of $m/z$ = 28 during irradiation is, therefore, an indication of CO photon-induced desorption.
\subsubsection{The photodesorption mechanisms}
\begin{figure}
\centering
\includegraphics[width=8.5cm]{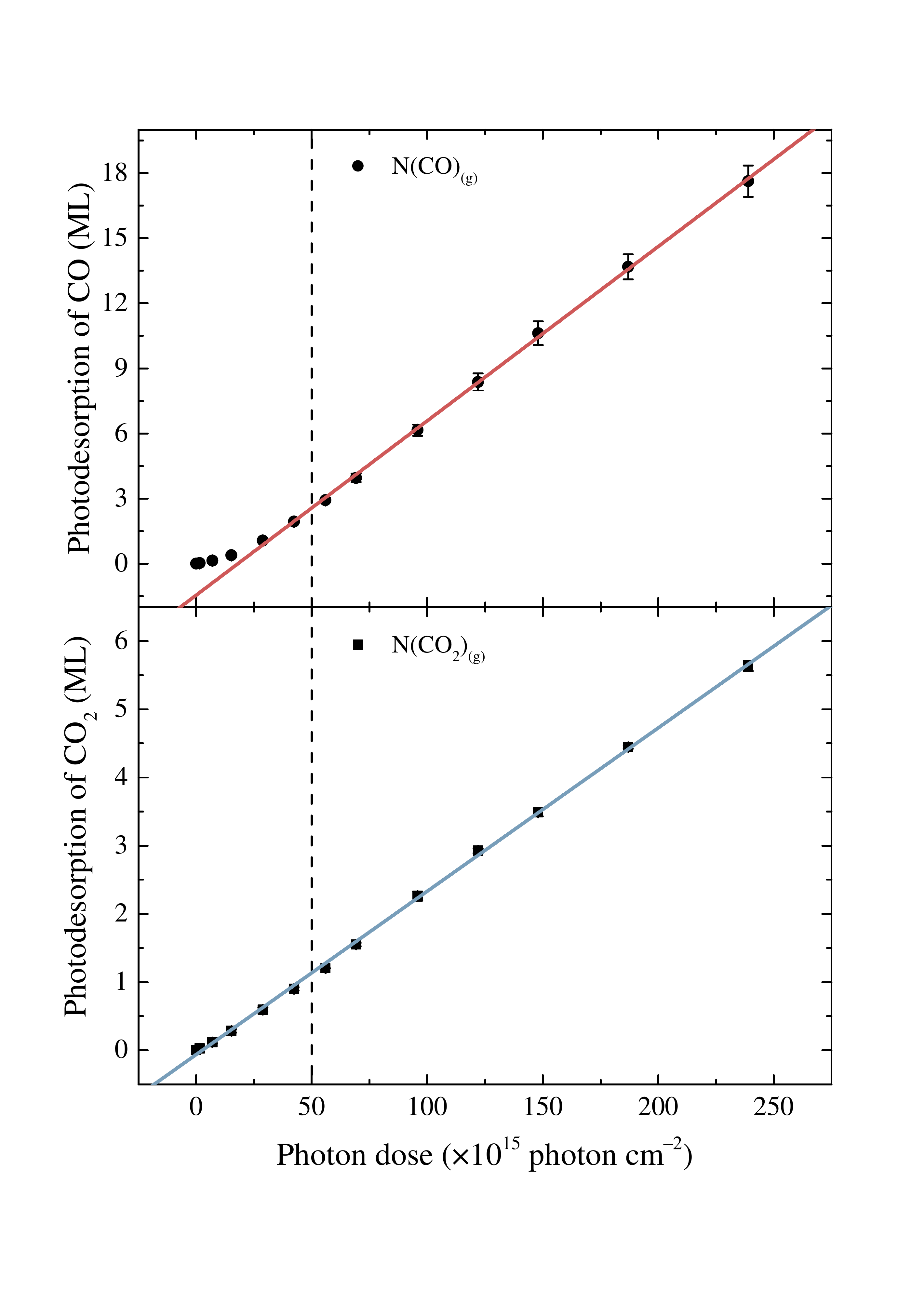}
\caption{The photodesorption of CO and \ce{CO2} as a function of photon dose in the 16 K deposition and irradiation experiment derived from calibrated QMS.}
\label{F4}
\end{figure}

There are basically two types of photodesorption mechanisms taking place during the irradiation of a pure \ce{CO2} ice (see also \citealt{fillion14}, \citealt{martin15}). One is photochemidesorption, which means that \ce{CO2} is photodissociated to form CO at the surface of \ce{CO2} ice with enough kinetic energy to desorb directly as a CO gas molecule at a decreasing rate with photon dose (concentration of parent molecules is reducing with photon dose). The other one is DIET, which occurs in the ice bulk, as explained in Section~\ref{sec:intro}, and dominates the photodesorption process when both processes take place simultaneously \citep{bertin12, fillion14, martin15}. The CO or \ce{CO2} molecules inside irradiated \ce{CO2} ice absorb a photon, and transfer its energy sequentially to the molecules near the surface layers to trigger the desorption, which is therefore an indirect photodesorption process. \\In the first five irradiation cycles, the ion current signal of $m/z$ = 28 is increasing with photon dose (Figure~\ref{F3}), since the number of CO molecules formed during VUV irradiated \ce{CO2} ice are not sufficient to reach the saturation point. This indicates that the CO molecules are photodesorbing through an indirect mechanism, because otherwise the photodesorption yield would be decreasing with photon dose, as explained above. In Figure~\ref{F4}, when photon dose reaches $5\times 10^{16}$ photons cm$^{-2}$, accumulated CO molecules reach 60\% of the column density of \ce{CO2} bulk, which is enough to support stably and efficiently the desorption yield via DIET mechanism. On the contrary, \ce{CO2} is the initial ice component and can desorb at constant rate during the whole irradiation sequence even when it is desorbing through the DIET mechanism, because the saturation point is already reached at the beginning of the experiment.
\subsection{The discrepancy in the \ce{CO2} photodesorption yield measured by infrared spectrometer and mass spectrometer} \label{sec:discrepancy}
\begin{figure}
\centering
\includegraphics[width=8.5cm]{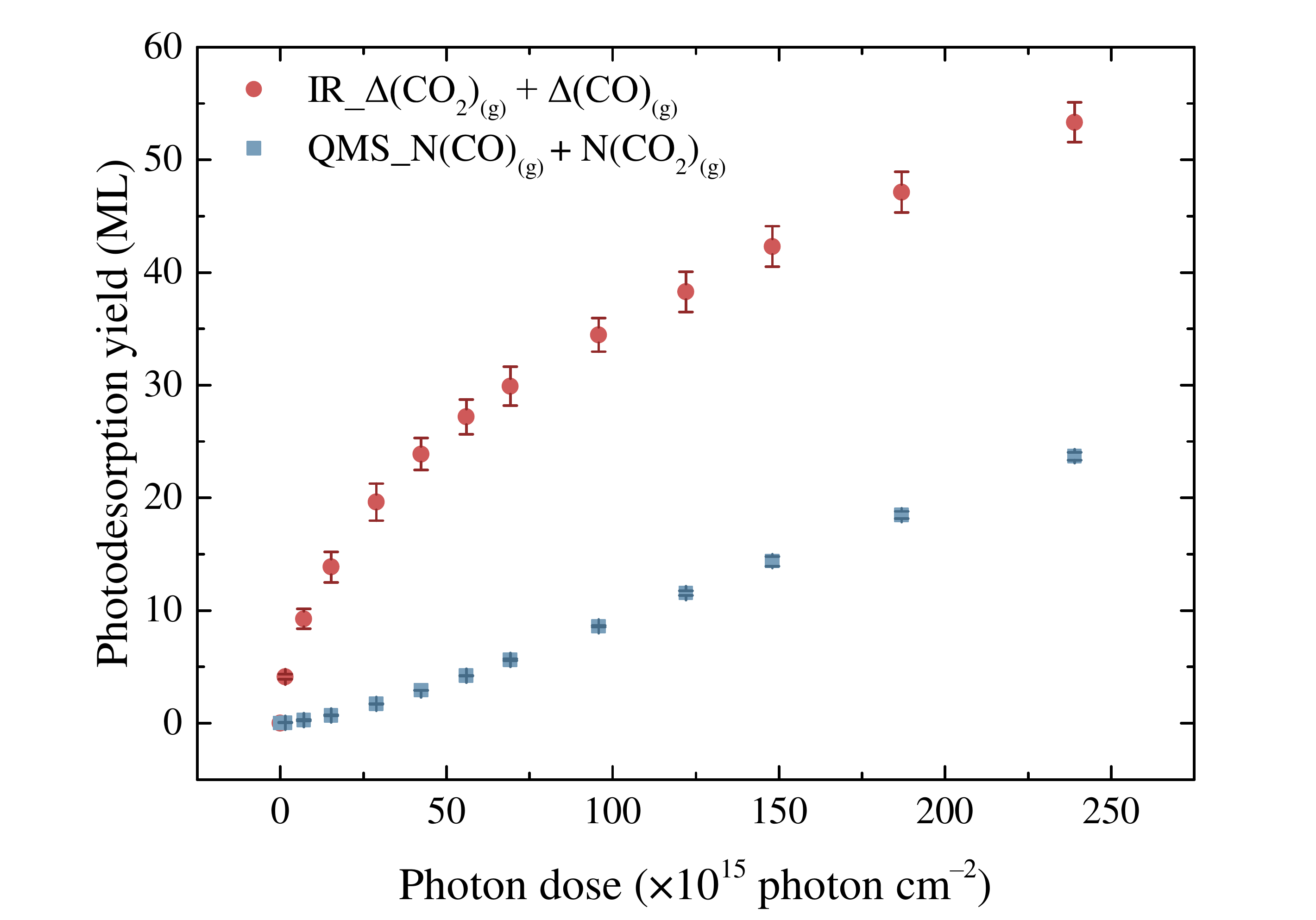}
\caption{The photodesorption yields derived from IR (carbon balance equation) and QMS as a function of photon dose in 16 K deposition configuration.}
\label{F5}
\end{figure}

In Figure~\ref{F5}, before photon dose reaches $2.5\times 10^{16}$ photons cm$^{-2}$, the IR data shows that the photodesorption rate increases very rapidly (sharp slope), and then decreases gradually, while the QMS data shows the opposite trend. Since the ion current from QMS represents a direct measurement of photodesorbing molecules in the gas phase, the photodesorption yield derived from QMS should be more reliable than the IR results; indeed, the later measures a decrease in the ice absorption band that can be due to other processes. We propose two possible reasons which may cause this discrepancy: missing carbon and changes in the ice IR absorption strength during irradiation. 
\subsubsection{Missing Carbon}
\begin{figure}
%\centering
\includegraphics[width=8.5cm]{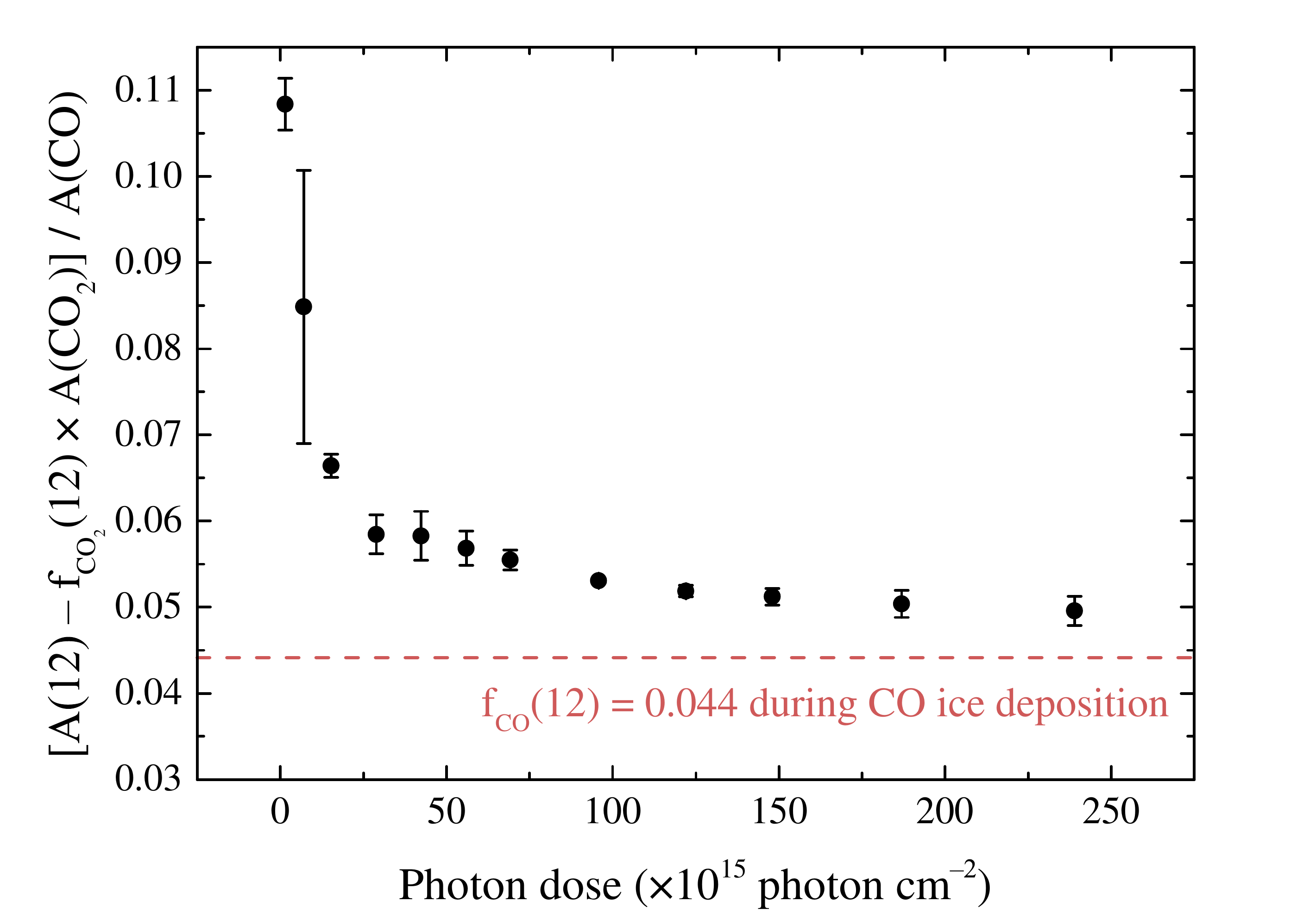}
\caption{The ratio of $\rm [A(12)-f_{\rm CO_2}(12)\times A(\rm CO_2)]/A(\rm CO)$ during VUV irradiation in experiment of \ce{CO2} deposited at 16 K configuration.}
\label{F6}
\end{figure}

\begin{figure}
%\centering
\includegraphics[width=8.5cm]{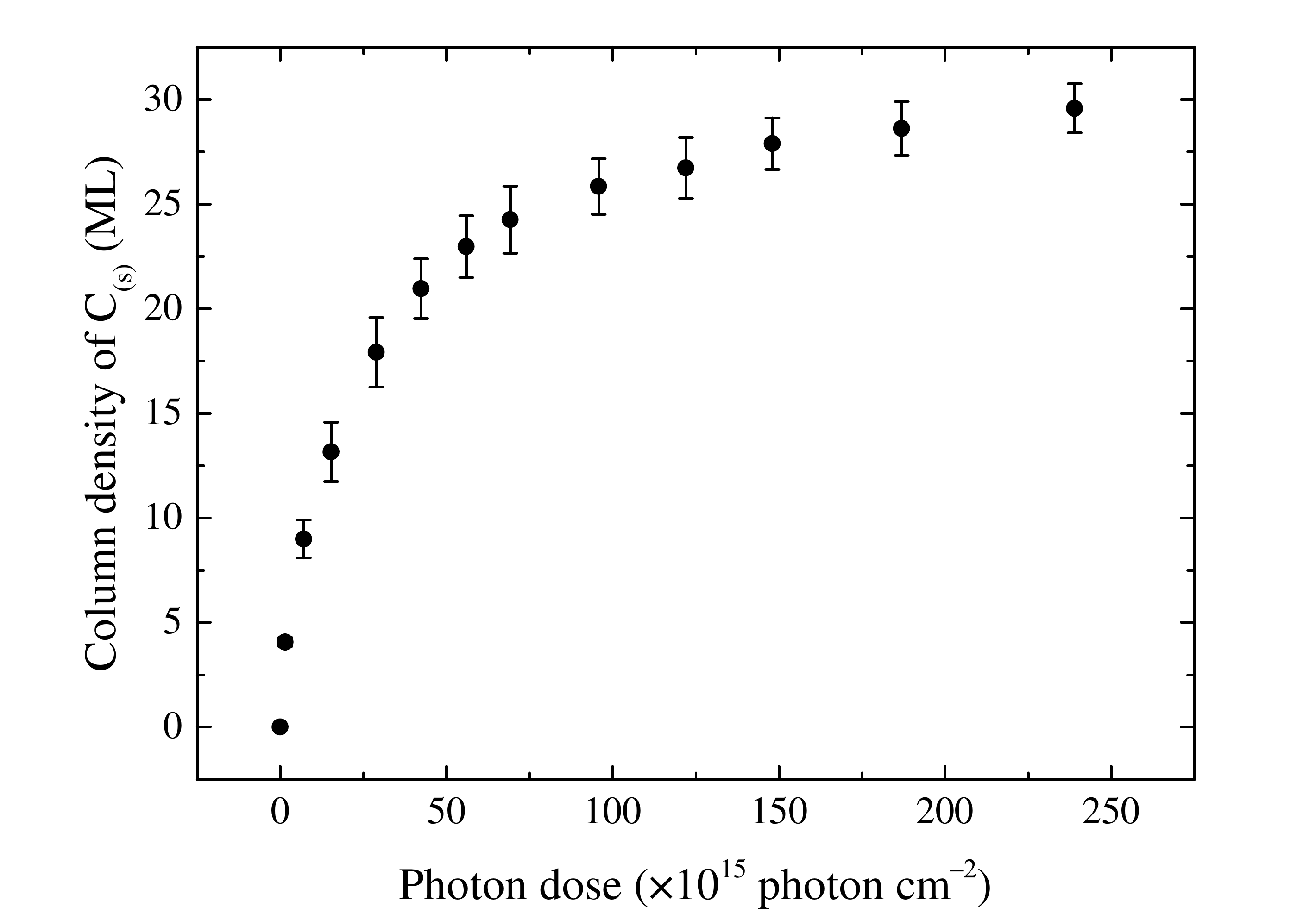}
\caption{The estimated column density of carbon atom as a function of photon dose in \ce{CO2} deposited at 16 K configuration.}
\label{F7}
\end{figure}

During VUV irradiation, \ce{CO2} dissociates as 
\begin{equation}\label{e5}
\rm CO_2+h\nu \to CO^*+O
\end{equation}
with the dissociation energy 5.45 eV of the C=O double bond \citep{darwent1970}, which is lower than the energy produced by MDHL (6.89--10.88 eV) to be a feasible reaction. Because the dissociation energy of the so-formed CO molecules (11.09 eV) is larger than the energy provided by the MDHL, one possible back reaction to reform \ce{CO2} is via the predissociation of CO \citep{gerakines96, loeffler05,G10,okabe78}: the excited CO* molecule combines with a ground state CO molecule through the reaction
\begin{equation}\label{e6}
\rm CO^*+CO \to CO_2+C,
\end{equation}
producing \ce{CO2} as well as a carbon atom, which is IR-inactive due to the lack of permanent dipole moment. Thus, when the photodesorption yields are calculated by carbon balance method, the carbon atom is not accounted for the calculation of Equation~\ref{e4}, which should be modified as
\begin{equation}\label{e7}
\begin{aligned}
\rm |\Delta CO_{2(s)}|=\Delta CO_{(s)}+\Delta CO_{3(s)}+\Delta C_{(s)}
\\\rm +\Delta CO_{(g)}+\Delta CO_{2(g)}+\Delta C_{(g)}
\end{aligned}
\end{equation}
\\Accordingly, considering a negligible formation of \ce{CO3} in the thin \ce{CO2} ice irradiation, the total number of \ce{CO2}, CO, and C photodesorbing species during the irradiation interval becomes $\rm |\Delta CO_{2(s)}|-\Delta CO_{(s)}-\Delta C_{(s)}$, which is lower than what we derived from carbon balance method, causing a discrepancy of results between IR and QMS. Figure~\ref{F6} shows evidence that supports the assumption of missing carbon: $\rm A(12)-f_{\rm CO_2}(12)\times A(\rm CO_2)$ includes the signal of photodesorption of C atom accompanied with the fragmentation of CO to C, $A(\rm C)/A(\rm CO)$. This ratio is referred to as $f_{\rm CO}(12)$ for pure CO gas injection into the UHV chamber. $f_{\rm CO_2}(12)$ is fragmentation of $A(\rm C)/A(\rm CO_2)$ during \ce{CO2} ice deposition. If the ratio of $\rm [A(12)-f_{\rm CO_2}(12)\times A(\rm CO_2)]/A(\rm CO)$ is larger than $f_{\rm CO}(12)$ during irradiation, it means there exists photodesorption of carbon. In the beginning, photodesorption of carbon atom is high compared to photodesorption of CO, which corresponds to the rapid desorption derived from carbon balance in IR measurement. After the photon dose reaches $1\times 10^{17}$ photons cm$^{-2}$, the photodesorption of carbon atoms is not so significant, and the amount of solid carbon atoms that remain in the \ce{CO2} ice bulk can be estimated from Equation~\ref{e7}, shown in Figure~\ref{F7}.

\subsubsection{Variation of absorption strength}
\begin{figure}
\centering
\includegraphics[width=8.5cm]{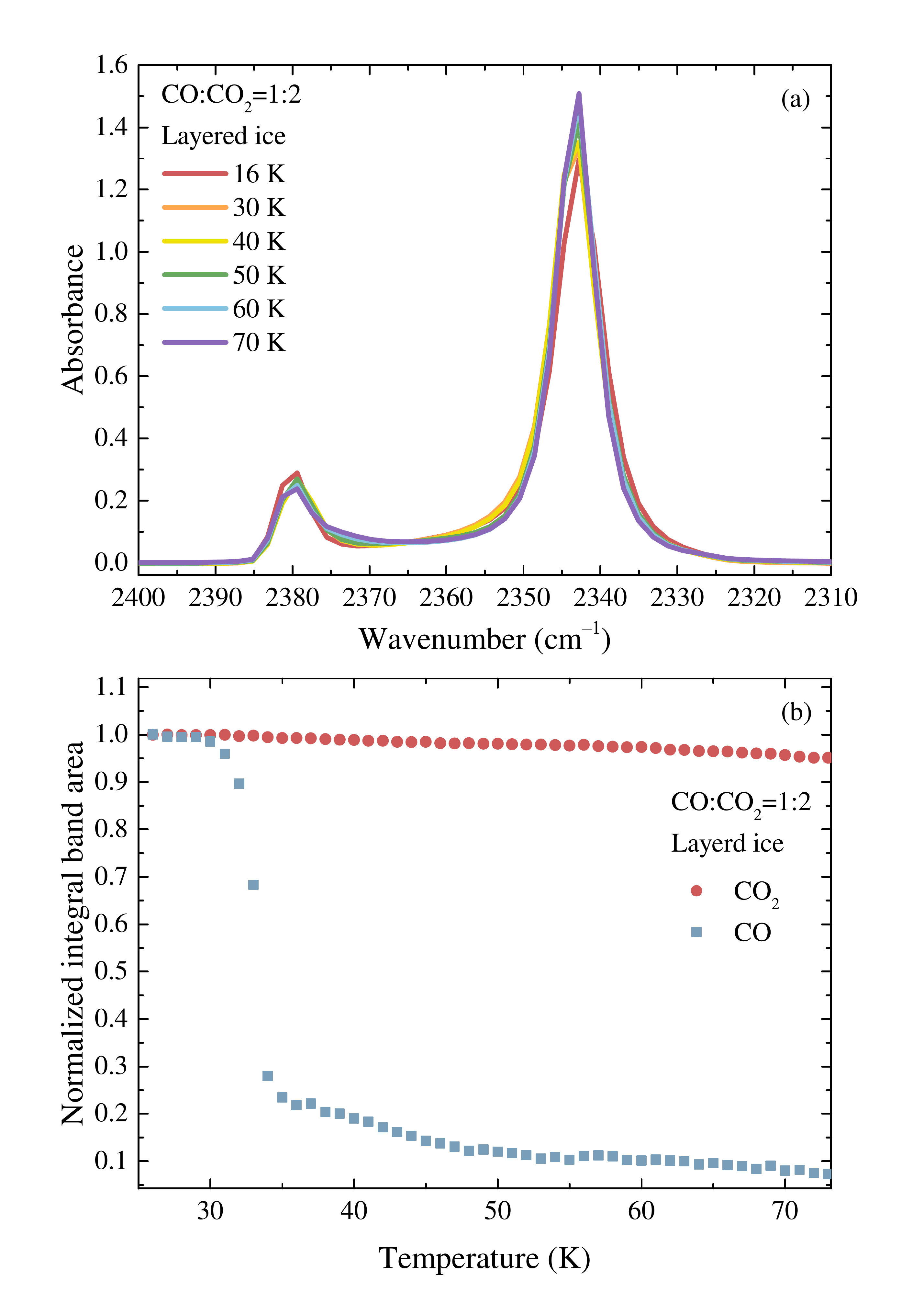}
\caption{(a) The absorbance of \ce{CO2} during annealing of the CO:\ce{CO2} = 1:2 layered ice. (b) The normalized integrated IR band of CO and \ce{CO2} during annealing of the CO:\ce{CO2} = 1:2 layered ice.}
\label{F8}
\end{figure}

\begin{figure}
\centering
\includegraphics[width=8.5cm]{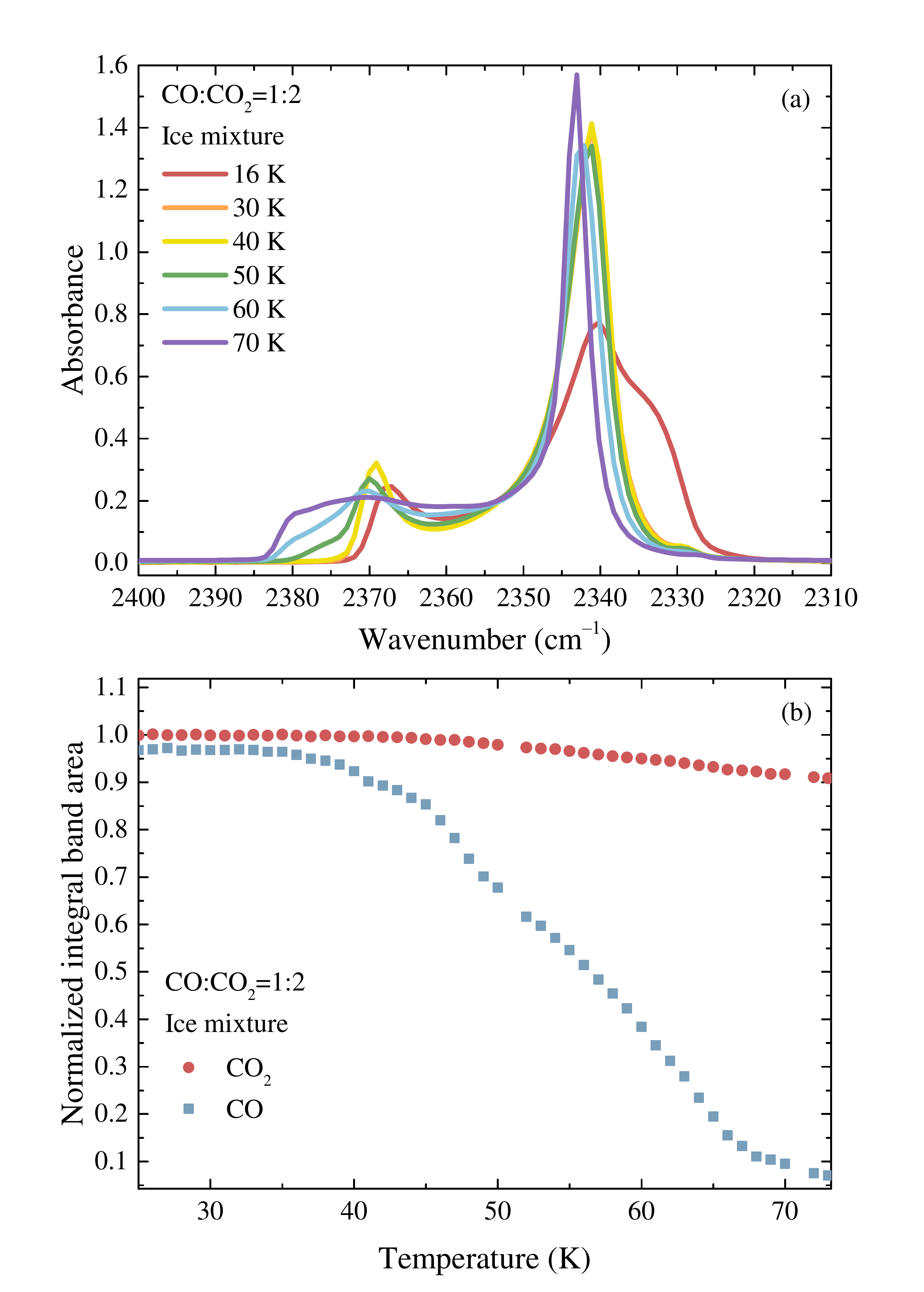}
\caption{(a) The absorbance of \ce{CO2} during annealing of the CO:\ce{CO2} = 1:2 ice mixture. (b) The normalized integrated IR band of CO and \ce{CO2} during annealing of the CO:\ce{CO2} = 1:2 ice mixture.}
\label{F9}
\end{figure}

The high concentration of CO molecules formed during irradiation in the \ce{CO2} ice bulk is expected to cause a change in the \ce{CO2} absorption band strength. Various authors reported changes in the IR band strengths (e.g., \citealt{d1982, oberg2007effects}), as a function of ice composition. We performed the following experiments to confirm whether the absorption strength of \ce{CO2} ice changed during irradiation. CO ice was deposited on substrate first, and then covered by \ce{CO2} ice on top at 16 K, with CO and \ce{CO2} column densities 74 ML and 139 ML (CO:\ce{CO2} = 1:2), respectively.  The layered ice was warmed up from 15 K to 95 K with heating rate of 0.5 K min$^{-1}$ for complete sublimation of \ce{CO2}. During annealing, CO started to thermally diffuse into \ce{CO2} ice through the pores, as described in \cite{cooke2018co}. The CO desorbed considerably at 30 K as shown in Figure~\ref{F8}, leaving \ce{CO2} behind. 
\\Furthermore, a mixture of CO and \ce{CO2} ice was deposited to monitor the abundant CO molecules in \ce{CO2} ice, with the similar column density and ratio as in layered ice experiment. During annealing, the CO molecules desorbed gradually from 40--70 K and then $\sim$5\% remained in the ice mixture, shown in Figure~\ref{F9}(b). The integrated absorption band of \ce{CO2} changed less than 10\% during thermal process, indicating that the variation of absorption strength of \ce{CO2} is not the main cause of discrepancy between the IR and QMS results, even though more than 60\% of CO molecules were produced in the \ce{CO2} ice bulk during irradiation.
\\According to the annealing experiments of \ce{CO2}/CO layered ice and \ce{CO2}+CO ice mixture, missing carbon is the most likely reason to cause the discrepancy of the trend of photodesorption yields derived from IR spectra and mass spectra. Therefore, we suggest that IR spectroscopy is not suitable for measuring the photodesorption yield of ices which can be easily photodissociated to form molecules or atoms without a dipole moment. 
\subsection{Photodesorption yield} \label{sec:Ypd}
\begin{deluxetable}{cccc}
\tabletypesize{\scriptsize} 
\tablecaption{Photodesorption Yields of CO, \ce{O2}, and \ce{CO2} Corresponding to Different Deposition Temperatures of \ce{CO2} Ices\label{t1}}
\tablewidth{40pt}
\tablehead{\\
  \ce{CO2} deposited temperature & $Y_{\rm pd}(\rm CO)$ & $Y_{\rm pd}(\rm O_2)$ & $Y_{\rm pd}(\rm CO_2)$  \\
 (K) & \multicolumn{3}{c}{$(\times 10^{-2}$ molecules photon$^{-1}$)}}
\startdata
16 & 8.3$\pm$0.4 & 2.4$\pm$0.1 & 2.4$\pm$0.2\\
30 & 7.6$\pm$0.4 & 2.3$\pm$0.1 & 2.$\pm$0.2\\
40 & 8.3$\pm$0.4 & 2.4$\pm$0.1 & 2.4$\pm$0.2\\
50 & 7.9$\pm$0.4 & 2.3$\pm$0.1 & 2.5$\pm$0.2\\
60 & 8.3$\pm$0.4 & 2.3$\pm$0.1 & 2.6$\pm$0.2\\
16$^a$ & 3.4$\pm$0.2 & 1.1$\pm$0.1 & 0.9$\pm$0.0\\
8$^b$ & $\sim$1.20 & $\sim$0.093 & $\sim0.011^c$\\
\enddata
\tablecomments{\\$\rm ^a$ Experiments with F-type MDHL lamp.\\$\rm ^b$ Adopted from \cite{martin15}.\\$\rm ^c~^{13}CO_2$}
\end{deluxetable}
Owing to the uncertainty of missing carbon in IR measurement and little variation of \ce{CO2} absorption band strength, the photodesorption yields ($Y_{\rm pd}$) at different deposition temperatures are calculated from the integrated ion current measured from QMS. Through the quantitative calibration of QMS, the photodesorption yield is no longer a relative value, but a meaningful photodesorption yield in units of molecules photon$^{-1}$. Average values of the photodesorption yield of CO, \ce{O2}, and \ce{CO2} under different deposition temperatures of \ce{CO2} ices are shown in Table~\ref{t1}. From Table~\ref{t1} it is obvious that \ce{CO2} morphology has no effect on photodesorption yield.

\begin{figure}
\centering
\includegraphics[width=8.5cm]{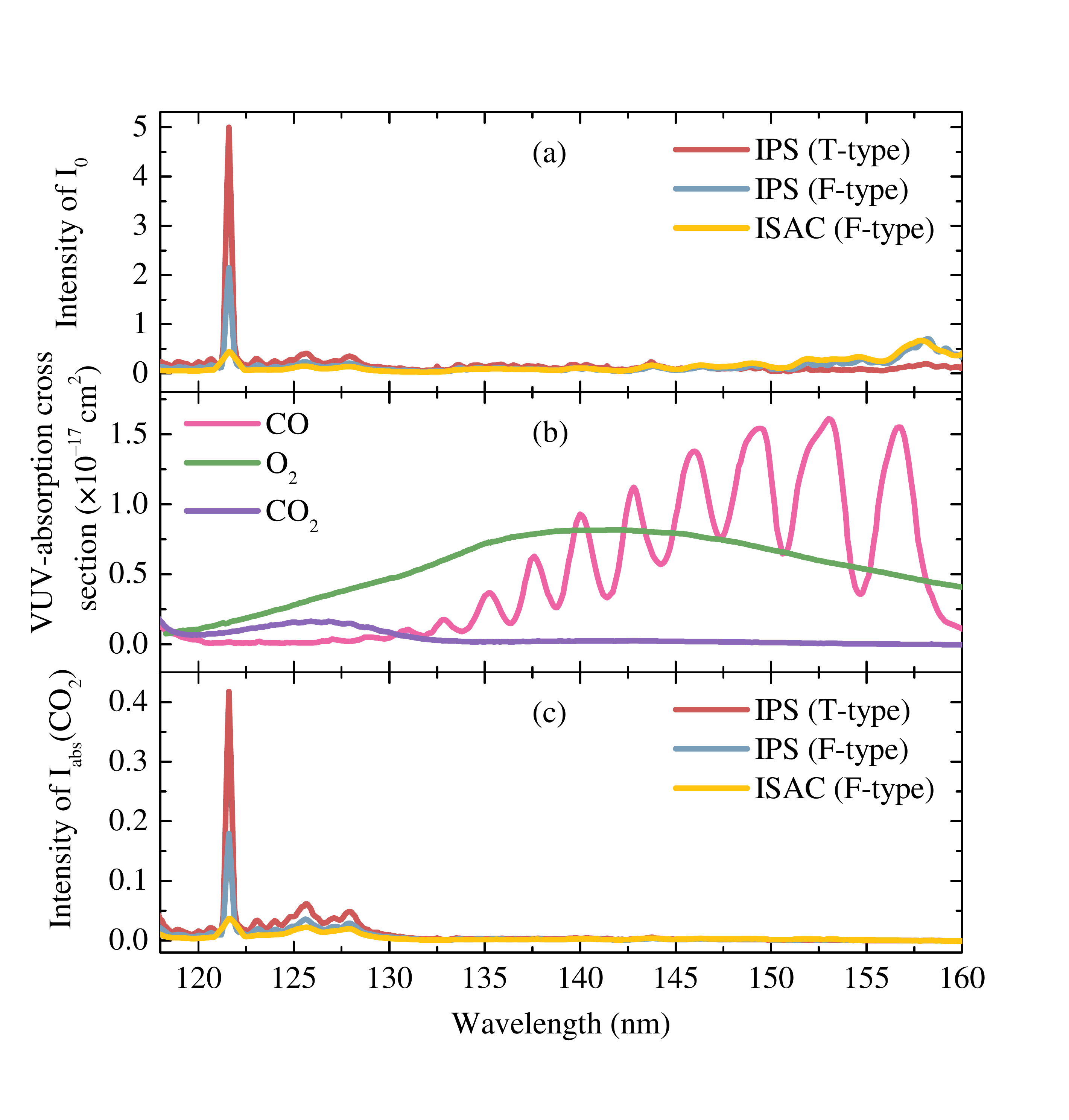}
\caption{(a) The incident intensity $I_0$ in three kinds of MDHL configurations. (b) VUV-absorption cross section of CO, \ce{O2}, and \ce{CO2}. (c) The weighted-absorption intensity of \ce{CO2} in three kinds of MDHL configurations.}
\label{F10}
\end{figure}

\begin{figure}
\centering
\includegraphics[width=8.5cm]{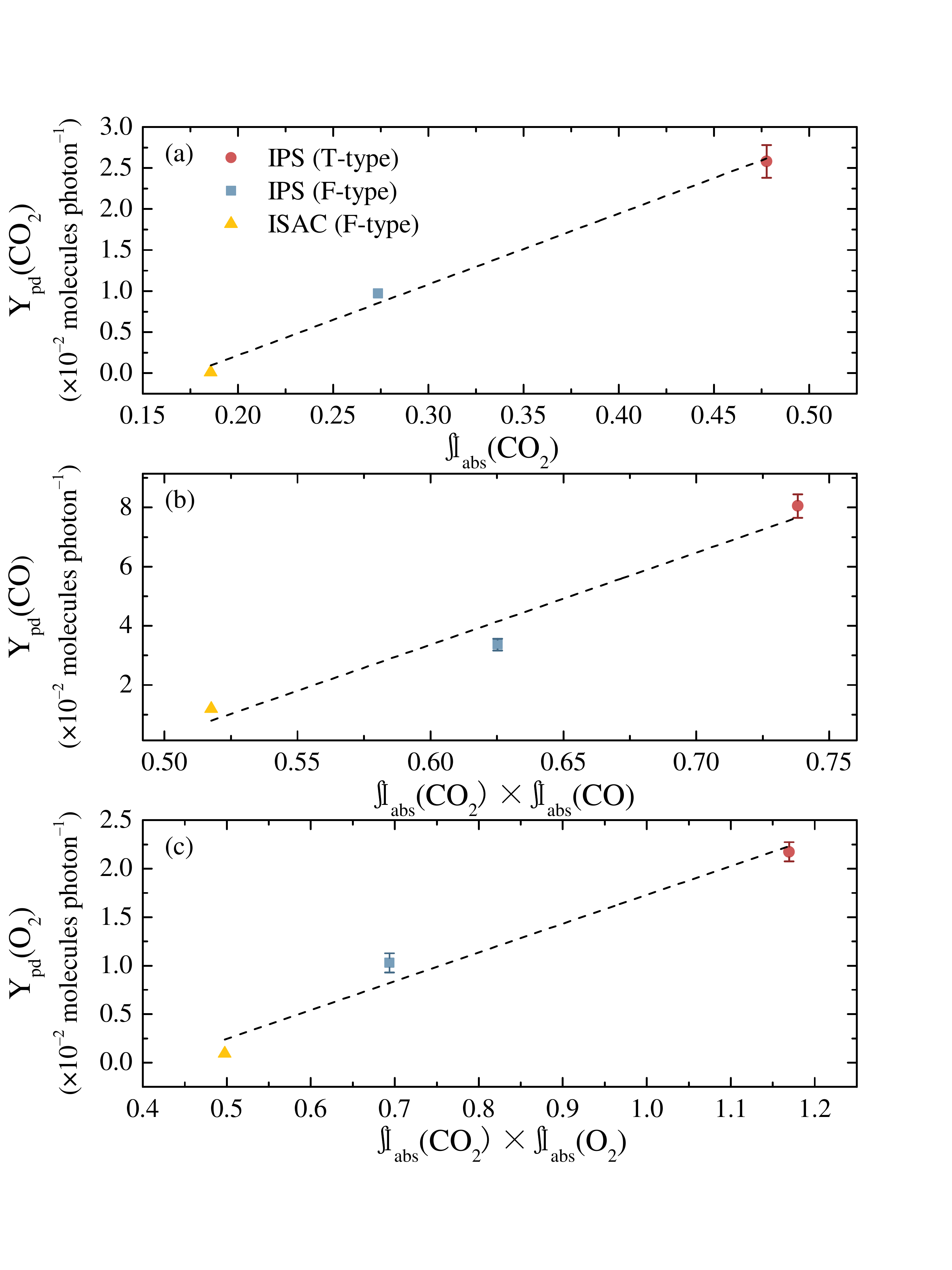}
\caption{The photodesorption yields of (a) \ce{CO2}, (b) CO, and (c) \ce{O2} as a function of (a) $\int I_{\rm abs}(\rm CO_2)$, (b) $\int I_{\rm abs}(\rm CO_2)\times I_{\rm abs}(\rm CO)$, (c) $\int I_{\rm abs}(\rm CO_2)\times I_{\rm abs}(\rm O_2)$ in three kinds of MDHL configurations with a linear fit. }
\label{F11}
\end{figure}

During VUV irradiation, the dominant photodesorbing species are CO and \ce{O2} molecules, in agreement with \cite{bahr2012photodesorption} and \cite{martin15}. Besides, the value we report for the observed \ce{CO2} photodesorption yield is significantly higher than previous estimates \citep{bahr2012photodesorption, fillion14, martin15}. We, therefore, employed $\rm ^{13}CO_2$ ice, deposited at 16 and 60 K, to discard a potential contamination of \ce{CO2} in the chamber, followed by irradiation at 16 K following the experimental procedure of $\rm ^{12}CO_2$ described in Section~\ref{sec:exppro} The ratios between photodesorption yields of CO/\ce{O2} and CO/\ce{CO2} at 16 and 60 K shown in Table~\ref{t2} are almost the same in both $\rm ^{12}CO_2$ and $\rm ^{13}CO_2$ cases, which corroborates that the photodesorption yield of \ce{CO2} is comparable to that of \ce{O2} in our experiments.
%未完成
\begin{deluxetable}{ccccc}
\tabletypesize{\scriptsize} 
\tablecaption{Ratios between Photodesorption Yields of CO/\ce{O2} and CO/\ce{CO2} in the Experiments of $\rm ^{12}CO_2$ and $\rm ^{13}CO_2$ Deposited at 16 and 60 K, Irradiating at 16 K\label{t2}}
\tablewidth{40pt}
\tablehead{\\
Deposited temperature & \multicolumn{2}{c}{16 K} & \multicolumn{2}{c}{60 K}  \\
\cline{2-3}
\cline{4-5}
$Y_{\rm pd}$ Ratio & CO/\ce{O2} & CO/\ce{CO2} & CO/\ce{O2} & CO/\ce{CO2}}
\startdata
$^{12}$C &3.39 &3.47 &3.58 &3.13\\
$^{13}$C &3.69 &3.47 &3.88 &2.96
\enddata
\end{deluxetable}
Photodesorption yield is dependent on the VUV-absorption cross section of the species and the VUV spectrum of MDHL. From Table~\ref{t1}, there is a large discrepancy of photodesorption yields of CO, \ce{O2}, and \ce{CO2} between \cite{martin15} and this work. \cite{chen14} reported that the geometry of quartz-tube of MDHL leads to different VUV spectra. We adopted a T-type lamp in IPS whereas a F-type lamp was used in ISAC \citep{martin15}. In addition, experiments using a F-type lamp were also performed in IPS. The photodesorption yields are shown in Table~\ref{t1}. According to Beer-Lambert law, the absorption spectrum of \ce{CO2} ice is
\begin{equation}\label{e8}
I_{abs}(\rm CO_2)=I_0-I_t=I_0(1-e^{-N\sigma_{CO_2}}),
\end{equation}
where $I_0$ and $I_t$ is the intensity of incidence and transmittance as a function of wavelength $\lambda$, N is the column density of \ce{CO2} ice, and $\sigma_{CO_2}$ the VUV-absorption cross section of \ce{CO2} as a function of wavelength $\lambda$. \\The normalized incident intensity in three kinds of MDHL configurations are shown in Figure~\ref{F10}(a), while (b) depicts the VUV-absorption cross section of CO, \ce{O2}, and \ce{CO2}. The weighted-absorption intensity of \ce{CO2} in three kinds of MDHL configurations derived from Equation~\ref{e8} are shown in Figure~\ref{F10}(c). The integration of CO\ce{CO2} absorption spectrum is positively correlated to the photodesorption, which is a zeroth-order process since it is not dependent on thickness in the thin ice experiments \citep{cottin2003photodestruction}. Figure~\ref{F11}(a) displays the photodesorption yields of \ce{CO2} as a function of integrated VUV absorption of \ce{CO2} convolved with the emission from the three kinds of MDHL configurations, which is represented by a linear relationship.\\CO is a photo-dissociation product from \ce{CO2} ice, and photon-induced desorption took place mainly through the DIET mechanism \citep{martin15}, therefore the photodesorption yield of CO is proportional to the product of integrated absorption spectrum of \ce{CO2} (production yield of CO molecules in \ce{CO2} ice) and integrated absorption spectrum of CO (energy transfer from excited CO molecules to those CO molecules in top few monolayers), shown in Figure~\ref{F11}(b). As for \ce{O2}, the formation is also a two-step reaction including Equation~\ref{e5} and
\begin{equation}\label{e9}
\rm O+O\to O_2.
\end{equation}
\\Analogous to the case of CO, the photodesorption of \ce{O2} is proportional to the multiplication of integrated absorption spectra of \ce{CO2} and that of \ce{O2}, shown in Figure~\ref{F11} (c).

\section{Conclusions and Astrophysical implications} \label{sec:con}
We have investigated the photodesorption yield of \ce{CO2} as a function of ice deposition temperature by transmittance FTIR and mass spectrometry. From IR spectral analysis, it is possible to estimate the total photodesorption yield of all the desorbing species, but information on the desorption of individual species cannot be extracted from this data alone. Moreover, we have demonstrated that the IR absorption strength of \ce{CO2} does not change significantly during irradiation. Therefore, the apparent discrepancy in the estimated photodesorption yields, obtained by IR and QMS, is explained when the IR-inactive carbon atoms formed by \ce{CO2} photoprocessing are taken into account. These carbon atoms reside in the \ce{CO2} ice bulk. With regard to the uncertainty in the photodesorption yield, we have performed a novel method for quantitative calibration of the mass spectrometer to study the photodesorption yield of \ce{CO2} ice deposited at different temperatures.\\The quantitative calibration of QMS is significant and indispensable to investigate the photodesorption yields. CO, \ce{O2}, and \ce{CO2} are the dominant desorbing species with the following photodesorption yields: $Y_{\rm pd}$(CO) = $8.3\pm 0.4\times 10^{-2}$ molecules photon$^{-1}$, $Y_{\rm pd}$(\ce{O2}) = $2.3\pm 0.1\times 10^{-2}$ molecules photon$^{-1}$, and $Y_{\rm pd}$(\ce{CO2}) = $2.4\pm 0.2\times 10^{-2}$ molecules photon$^{-1}$. The apparent discrepancy in the photodesorption yields between this work and \cite{martin15} can be explained very well by the linear relationship between the photodesorption yield and the integrated VUV-absorption in three kinds of MDHL configurations. In \cite{oberg09}, the $Y_{\rm pd}$(CO2) for \ce{CO2} deposited at 18–-30 K is $2.3\times 10^{-3}$ molecules photon$^{-1}$ calculated from $1.2\times 10^{-3}\times(1-e^{-x/2.9})+1.1\times 10^{-3}\times(1-e^{-x/4.6}) $, where x is the ice thickness, i.e., x = 141 ML to compare with this work. The $Y_{\rm pd}$(\ce{CO2}) is 40\% lower for \ce{CO2} deposited at 60 K and cooled down to 18 K than that for \ce{CO2} deposited at 18 K. This seems to verify that in the case of \ce{CO2} deposited at 40--60 K, the UV photon energy contributed to restructure the ice from crystalline to amorphous to make an effective exit channel for molecules to desorb to the surface, in line with the conclusion from \cite{yuan2014radiation}. In contrast, this study demonstrates that the photodesorption yield of \ce{CO2} is not ice structure dependent, which is incompatible with the preceding works based on IR spectroscopy for quantification \citep{oberg09, yuan2014radiation}. As mentioned in \cite{G16}, the photodesorption yield of CO ice is independent of ice structure, and the linearly decreasing photodesorption yield for increasing deposition temperature might be due to orientational disorder. Therefore, whether polar CO or non-polar \ce{CO2}, their structures are not directly related to the photodesorption yield.\\This work has implications for the interpretation of gas-phase observations toward cold interstellar and circumstellar environments where ice mantles were submitted to UV irradiation. The ice analog component studied in this paper, pure \ce{CO2} ice, is believed to form in ice mantles by two processes. One is the \ce{CO2} segregation out of the \ce{CO2}-\ce{H2O} mixture, leading to \ce{CO2} inclusions in water ice. The other is a distillation process, in which CO evaporates from a \ce{CO2}-CO ice mixture, leaving pure \ce{CO2} behind \citep{kim2012co2, pontoppidan2008c2d}. In the former process, UV irradiation of \ce{CO2} inclusions in water ice produces CO, \ce{O2} and \ce{CO3} molecules that will be mainly retained in the ice with a negligible desorption. The latter evolutionary phase of the ice is favorable for the photodesorption of \ce{CO2}, because \ce{CO2} ice is exposed to the surface of ice mantles, and our experimental results are more directly applicable to this scenario. Only \ce{CO3} is not expected to photodesorb in this case, according to the experiments. In particular, the photodesorption yields of \ce{CO2}, the CO and \ce{O2} photoproducts, in molecules photon$^{-1}$, should be valid to estimate the density of these species in the gas of cold environments. In dense clouds, the secondary UV field originated by excitation of molecular hydrogen is about $10^3-10^4$ photons cm$^{-2}$ s$^{-1}$, and should be dominated by the molecular emission lines with a smaller contribution of the more energetic Ly-$\alpha$ photons produced by atomic hydrogen. This UV emission spectrum, along with the total value of the UV flux, dictates the values of the photodesorption yields in our experiments. In Section~\ref{sec:Ypd} of this paper, we report the linear dependence of the photodesorption yields with the UV lamp emission and ice absorption spectra in the UV. \\Pure \ce{CO2} ice is expected to be in the crystalline form in ice mantles, since the band around 4.30 $\mu$m (2328 cm$^{-1}$) of amorphous \ce{CO2} ice is absent in the observed spectra toward dense clouds and protostars; this suggests that \ce{CO2} ice was formed on the dust at temperatures above 25 K, or was submitted later on to these temperatures \citep{escribano2013crystallization}. We confirmed that this ice structure reflects the temperature of ice deposition in the experiments. One of the conclusions from our work is that the photodesorption yield of \ce{CO2} does not depend on the degree of amorphous or crystalline structure of the ice.\\In Section~\ref{sec:discrepancy} of this paper, missing carbon is introduced to explain the \ce{CO2} discrepancy in the photodesorption yield measured by infrared spectrometer and mass spectrometer. When carbon atom is considered in the carbon balance method, the estimated ratio of $\rm \Delta C_{(s)}/\Delta CO_{2(s)}$ is $\sim$33\%. Unlike \ce{CO2}, other C-bearing molecules present in ice mantles, in particular CO, \ce{CH3OH}, and \ce{CH4}, are not expected to contribute appreciably to the release of C atoms in the gas phase. In the case of CO, only the most energetic photons with $E>$ 11 eV are capable to dissociate CO molecules to produce C. Meanwhile, photoprocessing of simple hydrocarbons in the ice, like \ce{CH3OH} or \ce{CH4}, trigger the formation of numerous species and no free C atoms are produced.\\A possible C-chemistry driven by secondary-UV in \ce{CO2}-rich ice will be a minor effect compared to X-rays, in particular in thin ices. But a considerable amount of C retained in the ice bulk, and close to the surface as in one of the ice scenarios (the one where CO distillates near 20 K and \ce{CO2} is the main component on the ice mantle surface), could have implications for the evolution of dense cloud interiors, and later on during star-formation. Indeed, the [C I] line was found to be a more important coolant than the lowest three rotational transitions of CO in moderate density gas, $n(\rm H_2)$ $\sim$a few $10^3$ cm$^{-3}$ as long as $N$(C)/$N$(CO) $\sim$0.1 \citep{pineau1992, wilson1997}. However, the ratios C/CO and C+/CO drop sharply when $\rm n_H$ approaches $10^4$ cm$^{-3}$ \citep{pineau1992}. Therefore, the background radiation field ionizes neutral C atoms, and C II is regarded to be an efficient coolant of the gas in star-forming regions within dense clouds (e.g. \citealt{stahler2008formation}).\\The release of C atoms to the gas phase might also have important implications for chemistry. At very low gas temperatures, neutral-neutral reactions are slow and C can stay longer in its atomic form, while at high temperatures, C will react with neutrals forming CO, HCN, \ce{C2}, etc.; see, for instance, the KIDA database (http://kida.obs.u-bordeaux1.fr/).\\Carbon-chain molecules account for $\sim$40\% of the interstellar molecular species detected to date, being specially abundant in dark clouds. The starless core Taurus Molecular Cloud-1 Cyanopolyyne Peak (TMC-1 CP) is the richest known source of carbon-chain molecules and is where most of the carbon-chain molecules detected in interstellar clouds exist (e.g., Fossé et al. 2001, Kaifu et al. 2004). Recently, carbon-chain molecules have also been detected in dense and warm regions around low-mass protostars. The carbon-chain molecules are thought to be produced from CH4 evaporated from dust grains via a mechanism called warm carbon-chain chemistry (WCCC), for which representative sources are L1527 and IRAS 15398-3359 (e.g., \citealt{sakai2008abundant}). The enhancement of carbon-chain molecules in these regions has been interpreted as the consequence of the evaporation of \ce{CH4} in a lukewarm (TK$\sim$50) region near a protostar \citep{hassel2008modeling, aikawa2008molecular}. The release of atomic C in this warm region might be an important ingredient for the chemical evolution of these objects.
\acknowledgments
This work has been supported by the MOST grants MOST 107-2112-M-008-016-MY3 (Y-JC), Taiwan, and project from the Spanish Ministry of Science, Innovation and Universities supported this research under grant number AYA2017-85322-R (AEI/FEDER, UE), and MDM-2017-0737 Unidad de Excelencia “María de Maeztu”-- Centro de Astrobiología (INTA-CSIC).
\bibliography{reference}

\end{document}